\documentclass[]{aa}  

\usepackage{graphicx}
\usepackage{txfonts}
\begin{document}

   \title{Collisions in Primordial Star Clusters}

   \subtitle{Formation Pathway for intermediate mass black holes}

   \author{B. Reinoso
          \inst{1}
          \and
          D.R.G. Schleicher\inst{1}
          \and
          M. Fellhauer\inst{1}
          \and
          R.S. Klessen\inst{2}$^{,}$\inst{3}
          \and 
          T. C. N. Boekholt\inst{4}
          }

   \institute{Departamento de Astronom\'ia, Facultad Ciencias F\'isicas y Matem\'aticas, Universidad de Concepci\'on, Av. Esteban Iturra s/n Barrio Universitario, Casilla 160-C, Concepci\'on, Chile; \\
              \email{breinoso@udec.cl}
         \and
             Universit\"at Heidelberg, Zentrum f\"ur Astronomie, Institut f\"ur Theoretische Astrophysik, Albert-Ueberle-Str. 2, 69120 Heidelberg, Germany
          \and
          Universit\"{a}t Heidelberg, Interdisziplin\"{a}res Zentrum f\"{u}r Wissenschaftliches Rechnen, Im Neuenheimer Feld 205, 69120 Heidelberg, Germany
          \and
             CIDMA, Departamento de F\'isica, Universidade de Aveiro, Campus de Santiago, 3810-193 Aveiro, Portugal
             } 

   \date{Received September 15, 1996; accepted March 16, 1997}

 
  \abstract{Collisions were suggested to potentially play a role in the formation of massive stars in present day clusters, and have
  likely been relevant during the formation of massive stars and intermediate mass black holes within the first star clusters. In 
  the early Universe, the first stellar clusters were particularly dense,  as fragmentation typically only occurred at densities 
  above $10^9$~cm$^{-3}$, and the radii of the protostars were enhanced due to the larger accretion rates, suggesting a potentially 
  more relevant role of stellar collisions. We present here a detailed parameter study to assess how the number of collisions as 
  well as the mass growth of the most massive object depends on the properties of the cluster, and we characterize the time 
  evolution with three effective parameters, the time when most collisions occur, the duration of the collisions period, as well 
  as the normalization required to obtain the total number of collisions.  We apply our results to typical Population III (Pop.~III) 
  clusters of about $1000$~M$_\odot$, finding that a moderate enhancement of the mass of the most massive star by a factor of a 
  few can be expected. For more massive Pop.~III clusters as expected in the first atomic cooling halos, we expect a more 
  significant enhancement by a factor of $15-32$. We therefore conclude that collisions in massive Pop.~III clusters were likely 
  relevant to form the first intermediate mass black holes.}


   \maketitle
%

\section{Introduction}
Collisions are often considered to be important during the formation of particularly massive stars, as 
suggested by \citet{Bonnell98} and \citet{Clarke08}. While the formation of low- and intermediate mass 
stars can be readily explained through accretion and infall in a protostellar core \citep[][see discussion 
by \citet{Palla93}]{Bodenheimer68, Larson69, Shu77}, 1D models originally failed for stars more massive 
than $20$~M$_\odot$, which start core hydrogen burning while still accreting, thereby preventing further 
accretion through ionizing radiation \citep{Yorke02}. More recent work shows however that in 3D situations, 
there are always channels through which accretion still occurs, thus not necesarrily providing a limiting 
factor \citep{Keto08, Krumholz09, Peters10a, Peters10b, Peters11}. A quantitative assessment of the role of
collisions in present-day protostellar star clusters has been pursued by \citet{Baumgardt11}, finding that
between $0.1-1\%$ of the protostars participate in such collisions within typical clusters, potentially providing
a relevant enhancement for sufficiently large numbers of stars.  Similar investigations have been pursued 
by \citet{Moeckel11}, \citet{Oh12} and \citet{Fuji13}.

In the early Universe, the conditions are potentially even more favorable for the formation of massive objects
via collisional processes. In the primordial gas, the cooling is less efficient and predominantly driven by small
fractions of molecular hydrogen \citep[see e.g.][]{Omukai05}. As shown via numerical simulations, the typical 
densities where fragmentation occurs are of the order $10^9$~cm$^{-3}$ or higher 
\citep[e.g.][]{Clark11a, Clark11b, Greif11, Greif12, Smith11, Smith12, Latif13}, leading to the formation of dense 
clusters with radii of $0.1$~pc or even smaller. Trace amounts of dust grains may even trigger fragmentation at 
still higher densities 
\citep[e.g.][]{Schneider03, Schneider06, Omukai08, Schneider12, Klessen12, Dopcke11, Dopcke13, Bovino16, Latif16}, 
providing ideal conditions for the formation of very dense clusters.

In addition to the compactness of the cluster, the protostellar radii are also enhanced by the more rapid 
accretion expected in primordial or low-metallicity gas, thereby increasing the overall cross section for 
collisions. Large protostellar radii of up to $300$~R$_\odot$ were calculated in stellar evolution models 
by \citet{Stahler86} or \cite{Omukai01, Omukai03}. It has been shown that variable protostellar accretion 
rates can enhance them further \citep{Smith12}. A strong increase of the radii also seems possible in the 
presence of particularly high accretion rates of $\sim0.1$~M$_\odot$~yr$^{-1}$ 
\citep{Hosokawa12, Hosokawa13, Schleicher13, Haemmerle17, Woods17}, implying $\sim500$~R$_\odot$ for a 
$10$~M$_\odot$ star and potentially more than $1000$~R$_\odot$ for a $100$~M$_\odot$ star.

It is thus conceivable that collisions may play some role in typical Population III (Pop.~III) clusters as 
expected to form in so-called minihalos with about $10^6$~M$_\odot$, as well as in the more massive atomic 
cooling halos with masses of the order $10^8$~M$_\odot$. These more massive halos are frequently considered 
as the birth places of intermediate mass black hole seeds. Collisional processes were indeed mentioned as an 
important pathway in the seminal paper by \citet{Rees84}, and subsequently taken into account for instance 
in semi-analytical models by \citet{Devecchi10, Devecchi12}, and \citet{Lupi2014}. In N-body simulations employing cosmological 
initial conditions, \citet{Katz15} and \citet{Sakurai17} have shown that black holes with masses of 
$\sim10^3$~M$_\odot$ can be formed.

Here, we present a systematic investigation on how the formation of very massive objects depends on the properties
of the cluster. Our numerical setup and the initial conditions are described in section~\ref{setup}, and our 
results are described in section~\ref{results}, including the time evolution of a typical cluster, the number 
of collisions and mass of the resulting object found under different conditions, as well as the time required to 
achieve such an enhancement. In section~\ref{application}, the results are applied to primordial clusters, both
in the context of minihalos and the larger atomic cooling halos. A final summary and discussion is given in 
section~\ref{discussion}.

\section{Simulation setup}\label{setup}
We present a set of $N$-body simulations of stellar collisions in compact star clusters. We have not included the effect of 
a gaseous potential or modeled the effect of the gas in explicit terms. To zero order, if the latter is dominating the potential, it will primarily increase the 
velocity dispersion and thereby decrease the crossing time of the cluster, allowing for a larger number of 
crossing times in a given physical time. In addition, there can be other gas-related effects that may enhance the 
number of collisions, which we will discuss in section~\ref{uncertainties}. The calculations here thus provide a conservative lower 
limit on the number of collisions.
To perform the calculations we use a modified version of 
NBODY6\footnote{Webpage NBODY6:\\ https://www.ast.cam.ac.uk/~sverre/web/pages/nbody.htm} \citep{Aarseth2000} to 
treat collisions, where we switch off the stellar evolution package and instead explicitly specify the stellar 
radii to perform a parameter study. NBODY6 is a fourth order Hermite integrator which includes a spatial hierarchy
to speed up the calculations: the Ahmad-Cohen scheme \citep{Ahmad1973}. It also includes routines to treat tidal
circularization which is believed to be the main mechanism from which binaries are formed in star clusters. Another 
important routine included is the Kustaanheimo-Stiefel regularization \citep{Kustaanheimo65}, an
algorithm to treat binaries and close two body encounters more accurately and faster.

\subsection{The star clusters}
We investigate how the number of collisions and the mass of the final object depends on the number of stars $N$ and 
the radii of the stars R$_{\rm star}$. We model a compact cluster in virial equilibrium consisting of equal mass stars 
with a total stellar mass of M$_{\rm cluster}=10^4$ M$_{\odot}$ using a Plummer distribution for the stars \citep{Plummer1911}
with a Plummer radius of R$_{pl}=0.077$ pc, implying a half-mass radius of R$_h=0.1$ pc. With this configuration 
the crossing time of the cluster is 0.022 Myr. We vary the number of stars $N=100,500,1000,5000$ and keep the mass 
of the cluster constant, therefore the initial masses of the stars M$_{\rm ini}$ depend on $N$ as 
M$_{\rm ini}=$ M$_{\rm cluster}/N$. For a fixed number of stars $N$ we vary the stellar radii 
R$_{\rm star}=20,50,100,200,500,1000,5000$~R$_{\odot}$.

\subsection{Stellar collisions}
The standard version of NBODY6 includes a routine to treat stellar collisions and can be used only when the stellar
evolution option is activated. The stellar evolution package included in the code is useful for metallicities ranging 
from Z=0.0001 up to Z=0.03 \citep{Hurley2000}. In this investigation, we are primarily interested in the application to
extremely metal poor conditions, with metallicities $Z\leq 10^{-6}$. Instead of pursuing detailed stellar evolution 
calculations, our goal here is to determine the regime where the collisions are potentially relevant. To treat the 
collisions, we therefore switch off the stellar evolution routine in NBODY6, and model the collisions as follows:

A collision occurs when the separation $d$ between two stars is smaller than the sum of the radii of the stars 
($d \leq R_1 + R_2$). When this condition is satisfied, we replace the two colliding stars by a new one, and the
mass of the new star M$_{\rm new}$ is simply the sum of the masses of the two colliding stars M$_{\rm new}= $ M$_1 + $M$_2$. 
The radius of the new star R$_{\rm new}$ is calculated with Eq.~\ref{eq:newradius} using the condition that the new star 
should have the same density as the colliding stars:
\begin{eqnarray}
\label{eq:newradius}
R_{\rm new} & = & R_1 \left( \frac{M_1 + M_2}{M_1}\right)^{1/3} .
\end{eqnarray}
Effectively, this assumes that the collision product quickly settles into a new equilibrium configuration in which 
the density corresponds to that of an unperturbed star of the same mass. This is consistent with detailed stellar evolution 
calculations, e.g., by \cite{Hosokawa12}  or \cite{Haemmerle16}. Nevertheless, we emphasize that we do not aim here to 
follow the detailed stellar evolution.

\section{Results}\label{results}
In the following, we describe the main results of our calculations. This includes the description of the time evolution 
in a typical cluster, the number of collisions that occur for different cluster parameters assuming a sufficient time
of integration, a description of the time evolution and its parametrization, a discussion of ejections as well as a 
discussion on uncertainties and neglected processes.

\subsection{Time evolution in a typical cluster}
\begin{figure}[h]
\includegraphics[width=250px]{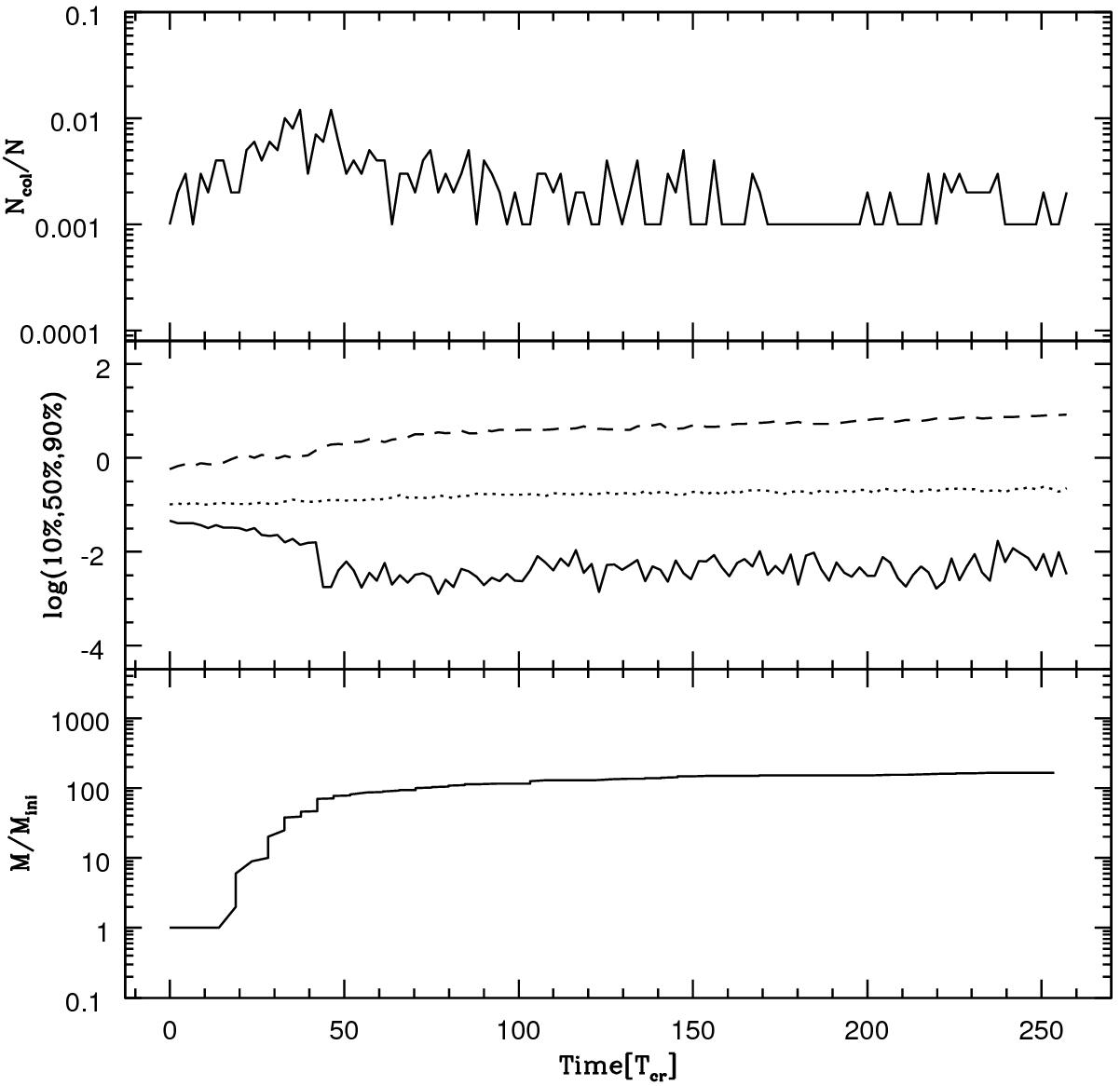}
\caption{Time evolution of a cluster with $N=1000$ stars, each of them with an initial radii of R$_{\rm star}=500$ R$_{\odot}$. 
Top panel: Number of collisions divided by number of stars as a function of time. Mid panel: Logarithm of the Lagrangian 
radii corresponding to $10\%$, $50\%$ and $90\%$ of the enclosed mass as a function of time. Bottom panel: Mass of 
the most massive object divided by the initial stellar mass as a function of time. In all panels, the time is normalized
by the cluster's crossing time.}
\label{evolutionN1000R500}
\end{figure}

\begin{figure}[h]
\includegraphics[width=250px]{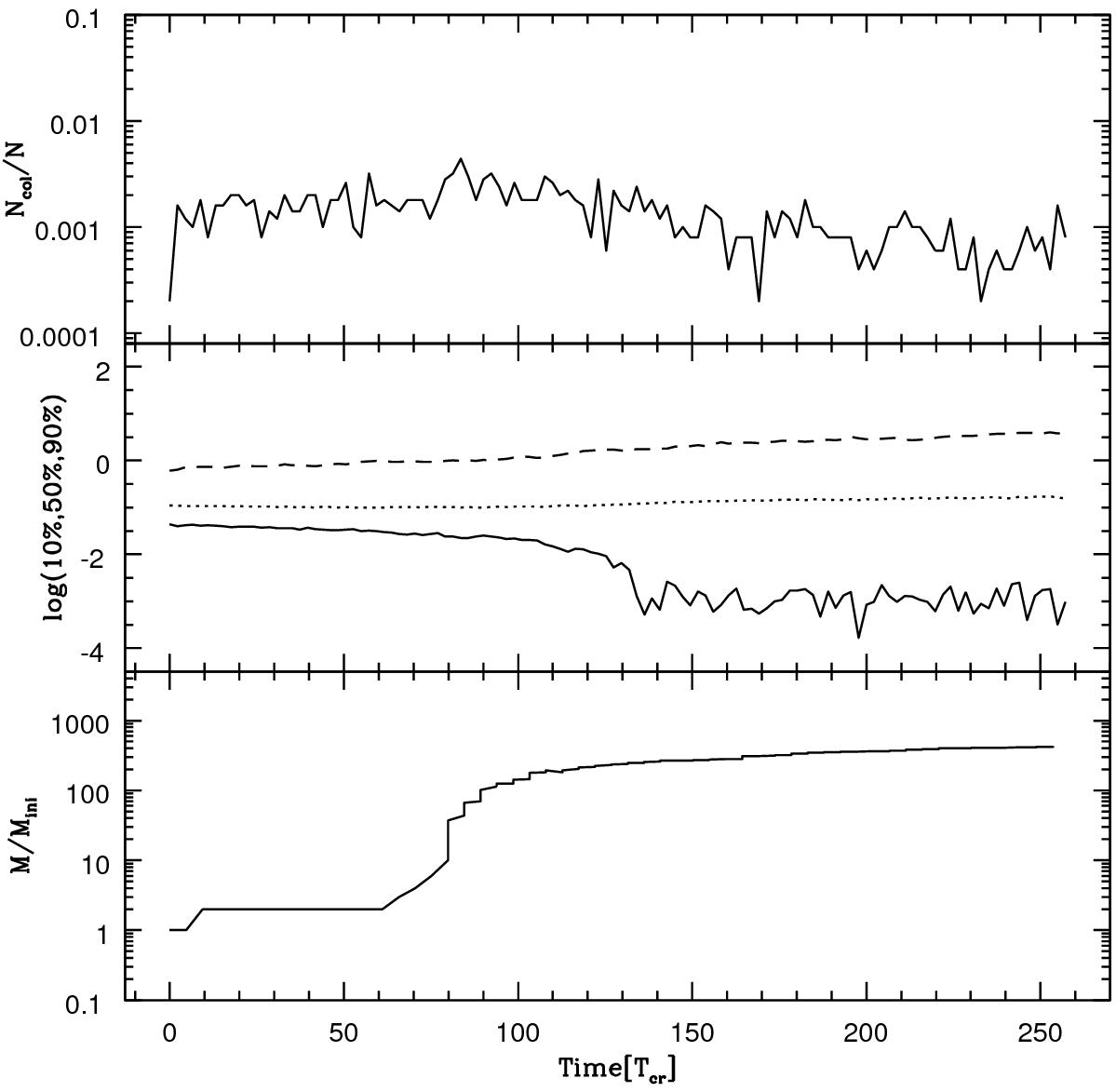}
\caption{Time evolution of a cluster with $N=5000$ stars, each of them with an initial radii of R$_{\rm star}=500$ R$_{\odot}$. 
Same panels as in Fig.~\ref{evolutionN1000R500}.}
\label{evolutionN5000R500}
\end{figure}

The time evolution in two typical clusters is shown in Figures ~\ref{evolutionN1000R500} and \ref{evolutionN5000R500}, 
the first one showing a cluster with $1000$ stars, and each of them having a radius of $500$~R$_\odot$. The second
one is a cluster with $5000$ stars and each of them having the same stellar radius (500 R$_{\odot}$). For both 
clusters, we show the fraction of stellar collisions (relative to the total initial number of stars), the logarithm of 
the Lagrangian radii corresponding to $10\%$, $50\%$ and $90\%$ of the mass, as well as the mass of the most massive
object normalized through the initial mass of the stars. 

In Fig.~\ref{evolutionN1000R500}, the maximum of collisions occurs clearly between 30-50 crossing times, where also the
Lagrangian radius corresponding to $10\%$ of the enclosed mass drops considerably, revealing that
a massive central object has already accreted $10\%$ of the mass at this time. Most of the collisions occur within the 
half mass radius of the cluster and more massive stars are formed very close to the center of the cluster where they 
collide to form a single very massive object. The peak fraction of collisions reaches about $1\%$, and the mass of the 
central object increases by a corresponding factor of about $100$, due to a similar number of stellar collisions. The
$50\%$ and $90\%$ Lagrangian radius shows sings of expansion, as the cluster dissolves due to ejections.

The behavior in Fig.~\ref{evolutionN5000R500} is quite similar. The main difference is that the time over which the 
collisions occur is more spread out, ranging from 10 up to $130-150$ crossing times. While the fraction of collisions 
is approximately constant during that period, with values around $0.1\%$, the run-away growth of the most massive 
object begins at around 80 crossing times, suggesting that before that stage, several more massive stars had formed 
which subsequently merge. 
Once the mass of the most massive object has increased by about a factor of $500$, thus corresponding to about $10\%$ of
the total mass, the corresponding Lagrangian radius decreases as seen in Fig.~\ref{evolutionN1000R500}. The overall 
evolution in both cases is thus quite similar.

\subsection{Number of collisions for different cluster parameters}
Our central goal is to determine how the number of collisions and the growth of the most massive object depends on 
the properties of the cluster. For this purpose, we will initially assume that enough time is available until all
collisions have occured, while investigating more details of the time evolution in the next subsection. 

\begin{table}
\begin{tabular}{|l|r|r|r|r|}
N & $B$ & $C$ & $D$ & $E$ \\
100 & 0.45 $\pm 0.02$ &-2.25 $\pm 0.06$  & 0.40$\pm 0.04$ &-0.16$\pm0.09$  \\
500 & 0.49 $\pm 0.02$ & -2.27 $\pm 0.06$ & 0.53$\pm 0.04$  & 0.22$\pm 0.09$  \\
1000 & 0.51$\pm 0.04$ & -2.29 $\pm 0.09$ & 0.57$\pm 0.06$   &0.44$\pm 0.14$  \\
5000 & 0.50$\pm 0.03$ & -2.16 $\pm 0.07$ & 0.52$\pm 0.05$  & 1.28$\pm 0.12$ \\
\end{tabular}
\vspace*{0.3cm}
\caption{Parameters from the fit to the functions Eq.~\ref{eq:fit1} (Column 2 \& 3) and Eq.~\ref{eq:fit2} (Column 4 \& 5) 
to estimate the total fraction of collisions in a cluster depending on the initial radii of the stars R$_{\rm star}$ and 
the initial number $N$ of stars (Eq.~\ref{eq:fit1}). Parameters $D$ \& $E$ are used to estimate the mass of the most massive 
object divided by it's initial mass (M$_{\rm max}$/M$_{\rm ini}$)at the end of the runaway growth depending on the initial radii 
R$_{\rm star}$ of the stars and the initial number $N$ of stars (Eq.~\ref{eq:fit2}).}
\label{fitmass}
\end{table}

\begin{figure}[h]
\includegraphics[width=250px]{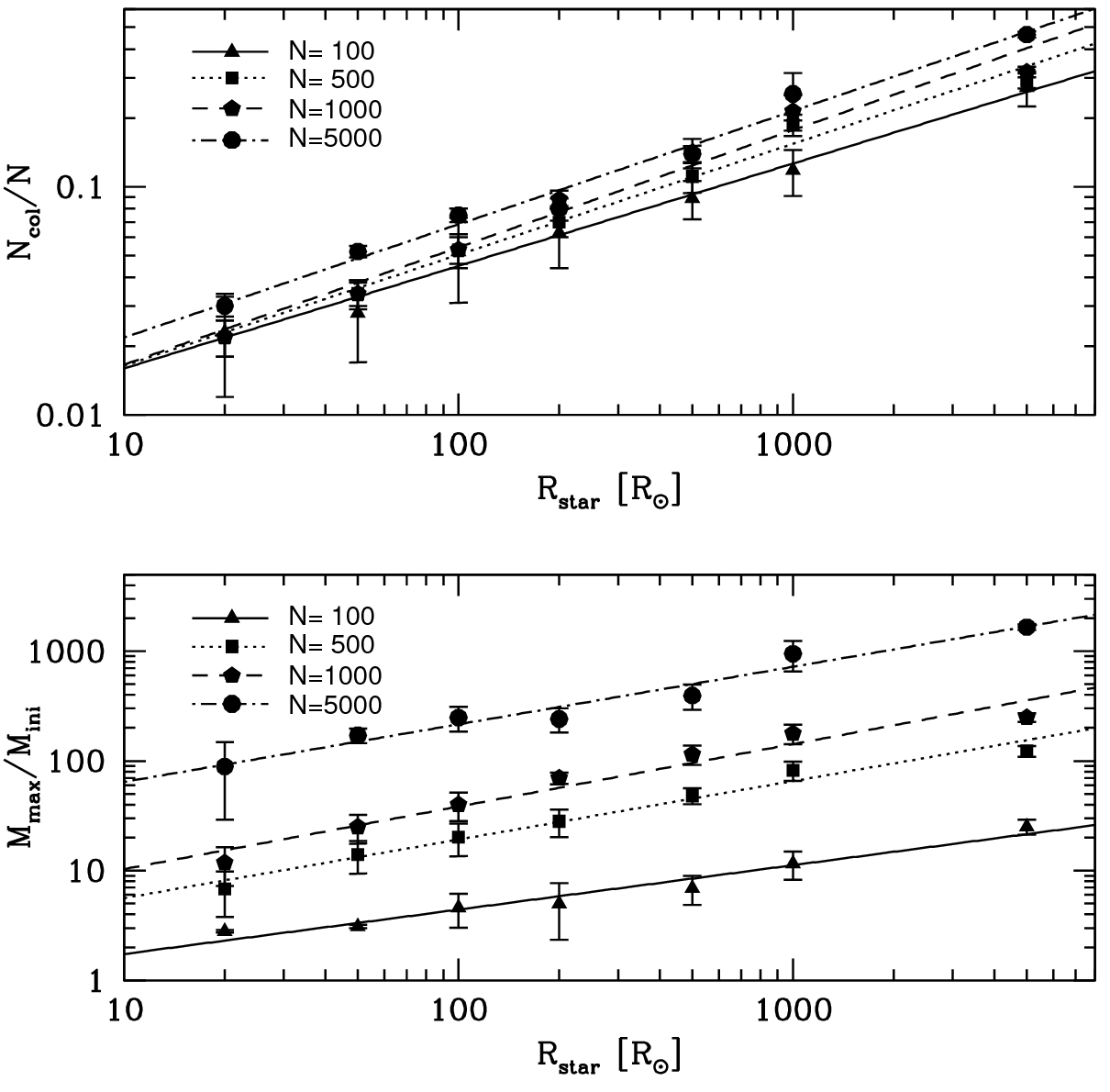}
\caption{Top Panel: Fraction of collisions (N$_{\rm col}/N$) as function of the initial stellar radius and the number of stars, 
we also show the best linear fits from Eq.~\ref{eq:fit1} and the parameters for these fits in Table~\ref{fitmass}.
Bottom Panel: Final mass over the initial mass (M$_{\rm max}/$M$_{\rm ini}$) for the most massive star at the end of 
the runaway growth as function of the initial stellar radius and the initial number of stars. We also show the best linear fits 
from Eq.~\ref{eq:fit2} and the parameters for these fits in Table~\ref{fitmass}.}
\label{NcolMassFit}
\end{figure}

Our main results are given in Fig.~\ref{NcolMassFit}, showing the fraction of collisions as a function of stellar 
radius and for different numbers of stars. The data shown here correspond to an average over $10$ simulations 
for each individual configuration, to improve the statistical reliability of the results. We find that the total
fraction of collisions depends very weakly on the number of stars, showing an increase by at most a factor 
of 2 when going from 100 to 5000 stars. This very minor increase may result from the fact that a larger 
(but still small) fraction of the effective area is filled with stars. We do in principle not expect this to depend 
strongly on the IMF, but it remains one of the uncertainties to be explored in future studies. The dependence on the 
stellar radii on the other hand is very clear and corresponds to a power-law with slope of about $0.5$ 
(see second column in Table~\ref{fitmass}).

The mass of the most massive object normalized by the initial mass of the stars shows a clearer dependence both on
number of stars and on radius. This can be understood from the fact that the fraction of collisions is roughly
independent of the number of stars. If most collisions occur with the most massive object (as we have checked), 
the fraction of the mass going into the most massive object is approximately constant, and thus normalizing it by
the initial mass of the stars leads to a dependence on the number of stars, as the cluster mass is fixed. As a 
function of stellar radius, this quantity indeed shows a power-law behavior with a slope of about $1/3$.
We fit the data using an implementation of the nonlinear least-squares Marquardt-Levenberg algorithm in \texttt{gnuplot}
with the functions:
\begin{eqnarray}
\label{eq:fit1}
 \log{\left( \frac{\rm N_{\rm col}}{N}\right) } \ &=&B \log(R_{\rm star}) + C , \\
\label{eq:fit2}
\rm  \log{\left( \frac{M_{\rm max}}{M_{\rm ini}}\right) } &=&D \log(R_{\rm star}) + E .
\end{eqnarray}

The parameters of the fit are shown in Table~\ref{fitmass}.
We found that $E$, $C$ does not depend on $N$, however even if this parameters are constant, the total 
number of collisions N$_{\rm col}$ still depends on $N$ as described in Eq.~\ref{eq:fit1}. We also found 
that $D$ is constant but $E$ depends on $N$ as follows:

\begin{eqnarray}
\label{eq:EonN}
E &=& 0.84 \pm 0.13 \log(N) - 1.96 \pm 0.39 .
\end{eqnarray}

\subsection{Description of the time evolution}
For many practical applications, it will be necessary to have more information on the time evolution of 
the collisions. While a detailed description of every run is clearly unfeasible, our goal is to extract 
three main parameters that contain the most essential information. This is the time delay when most of 
the collisions occur, the duration of this period, as well as the total number of collisions.  For this
purpose, a Gaussian fit is applied to the number of collisions over time. The parameter $t_{\rm delay}$
describes the time when most collisions occur, $t_{\rm duration}$ approximates the duration of the 
collision period. The fit is given as:\begin{equation}
N_{\rm col}=A \mathrm{exp}\left(-\frac{(t-t_{\rm delay})^2}{2t_{\rm duration}^2}  \right) .
\end{equation}
The normalization A is adopted to ensure that the total number of collisions is correctly reproduced.

\begin{figure}[h]
\includegraphics[width=250px]{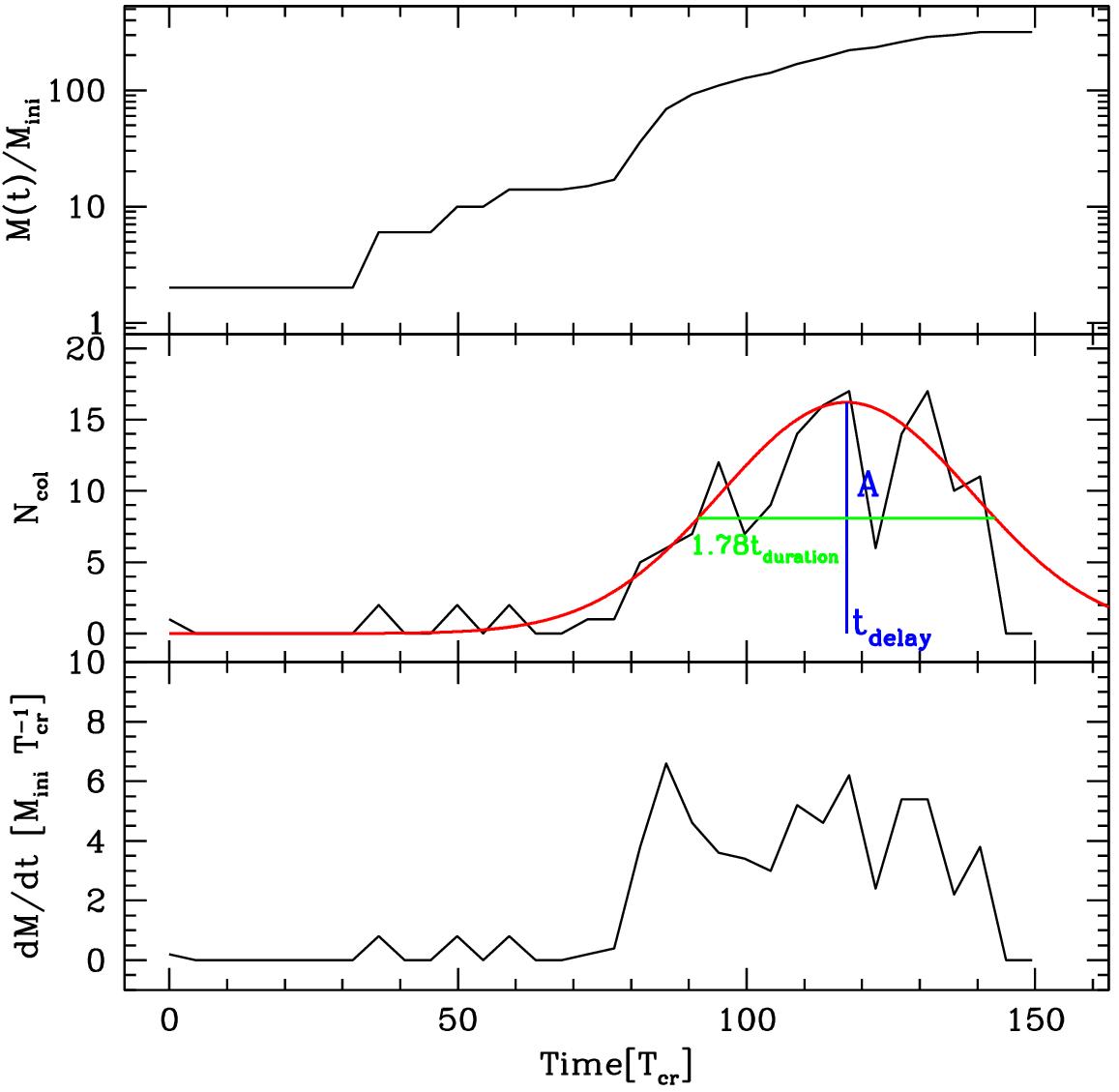}
\caption{Top Panel: Mass evolution of the runaway star (M$_{\rm max}(t)/$M$_{\rm ini}$) in a cluster 
with $N=5000$ stars, each of them with an initial mass of M$_{\rm star}= 2$M$_{\odot}$ and an initial radius of 
R$_{\rm star}=200$ R$_{\odot}$. Middle panel: Number of collisions in bins of 5 crossing times (N$_{\rm col}$) 
as function of the cluster's crossing time. The red line is the best gaussian fit. Bottom panel: Mass growth
rate $\rm dM/dt$ in M$_{\rm ini}\,T_{\rm cr}^{-1}$ calculated in bins of 5 crossing times.}
\label{fitgausian}
\end{figure}

An example for such a fit is given in Fig.~\ref{fitgausian}, showing the number of collisions in a cluster
with 5000 stars and a stellar radius of 500~R$_\odot$ as a function of time. The results shown in Fig.~\ref{fitgausian} 
are calculated in bins of 5 crossing times as there are few collisions per crossing time and we aim for a general descriprion 
of the effect of collisions rather than a more individual description that depends more on statistical variatons. Nevertheless
a smaller or larger binsize does not change our results as we have checked. While the 
time evolution of the collisions does not precisely follow a Gaussian distribution, we find that the time when most collisions 
occur as well as the duration of the collisions is described quite well by the Gaussian fit. For comparison, 
we also show the time evolution of the mass of the most massive object divided by the initial mass as a 
function of time.
The steepest growth in mass occurs around 80 crossing times due to a collision with a 40 M$_{\odot}$ star
formed also through stellar collisions, however, the number of collisions peaks around 120 crossing times, in
agreement with a second peak in the mass enhancement, which is produced at this stage by several mergers with smaller stars.
The width of the Gaussian fit matches about the time when the central object stops growing in mass. Given 
the overall uncertainties in the problem considered, the latter provides a reasonable description of 
the most relevant information.

\begin{figure}[h]
\includegraphics[width=250px]{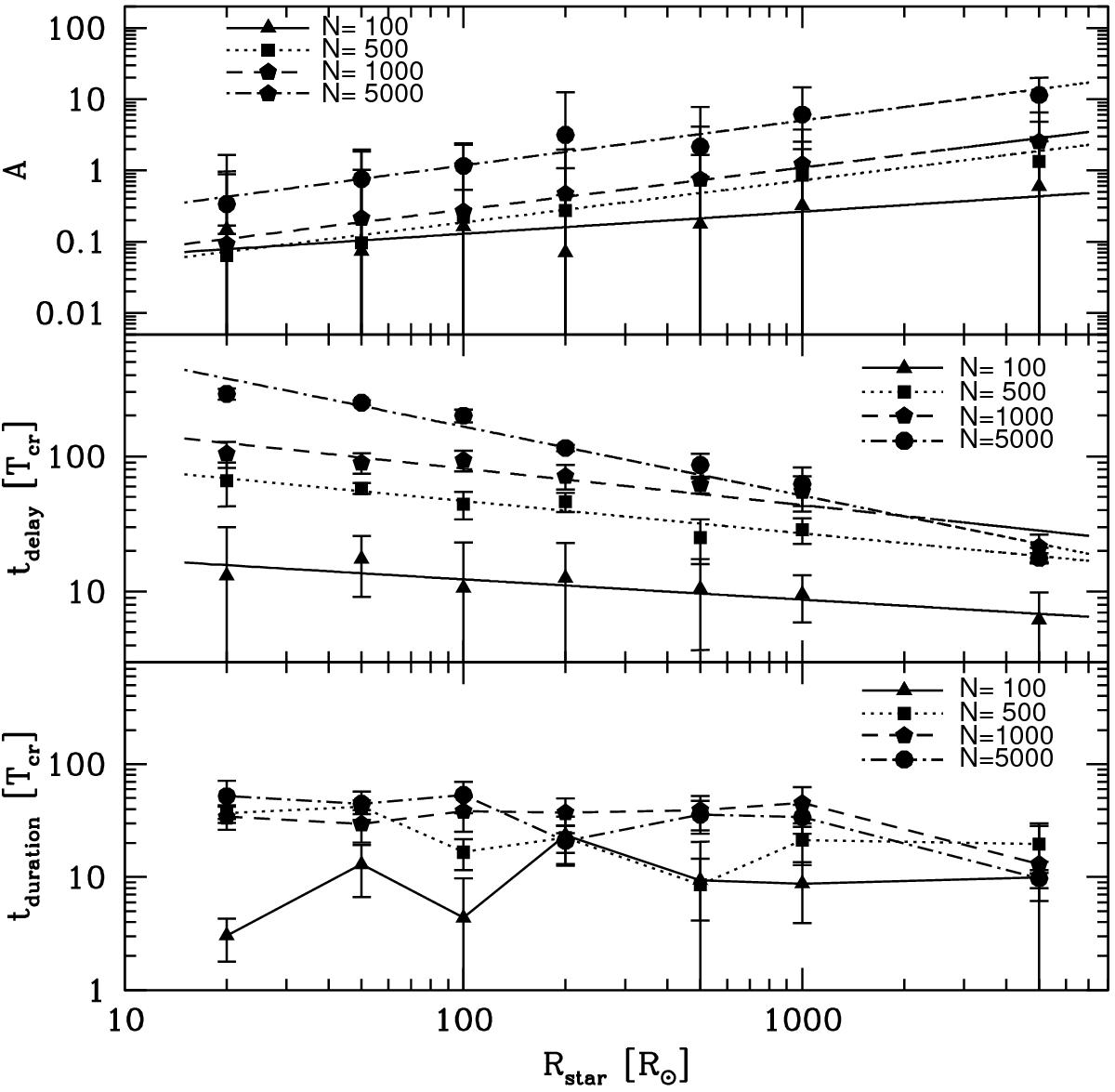}
\caption{Parameters of the gaussian fit A, $t_{\rm delay}$ and $t_{\rm duration}$ as function of the 
initial stellar radii (R$_{\rm star}$) and the initial number of stars $N$.}
\label{gaussianparameters}
\end{figure}

The resulting fit parameters are given in Fig.~\ref{gaussianparameters} as a function of stellar
radius and for different numbers of stars. The parameter $A$ increases by a factor of $4$ between
$20$ and $5000$~$R_\odot$ for a cluster with 100 stars, while the increase is more enhanced, corresponding
to about a factor of 20, for a cluster with $1000$ stars. The increase of the parameter $A$ is
more moderate than naively expected from the behavior of the total fraction of collisions, which is
partly due to the way how the duration of the collision period changes with the number of stars.
In particular, the evolution of the $t_{\rm duration}$ parameter as a function of stellar radius is 
relatively flat (slightly decreasing), and only weakly dependent on the number of stars. However, a
decreasing value of $t_{\rm delay}$, which quadratically enters the exponential, can overcompensate 
for the behavior of $A$. The parameter $t_{\rm delay}$ describing the time when most of the collisions 
occur, on the other hand, decreases with stellar radius and increases with number of stars. This reflects
in particular the increase of the relaxation time in the cluster with the number of stars, as well as 
the increased probability for collisions with increasing stellar radius. We fit a linear function to 
the quantities $\log(A)$ and $\log(t_{\rm delay})$, with $t_{\rm delay}$ expressed in units of the 
cluster's crossing time $T_{cr}$ depending on the logarithm of the stellar radii
as follows:\\

\begin{eqnarray}
\label{eq:parameterA}
 \log(A) &= & \alpha \log(\rm R_{\rm star}) + \beta , \\
\label{eq:parameterb}
 \log(t_{\rm delay}) &= & \gamma \log(\rm R_{\rm star}) + \delta .
\end{eqnarray}

The parameters $\alpha$, $\beta$, $\gamma$ and $\delta$ are shown as function of the number of stars 
$N$ in Table~\ref{table2} and the fitting lines are shown in Fig.~\ref{gaussianparameters}.

\begin{table}
\begin{tabular}{|l|r|r|r|r|}
N & $\alpha$ & $\beta$ & $\gamma$ & $\delta$ \\
100 & 0.31$\pm0.12$  &-1.51$\pm0.29$  & -0.15$\pm0.04$  & 1.39$\pm0.09$  \\
500 & 0.59$\pm0.06$ & -1.91$\pm0.16$  & -0.24$\pm0.03$  & 2.15$\pm0.07$  \\
1000 & 0.63$\pm0.07$  & -1.19$\pm0.19$  & -0.27$\pm0.05$   &2.45$\pm0.11$  \\
5000 & 0.59$\pm0.03$  & -1.73$\pm0.07$  & -0.51$\pm0.05$  & 3.24$\pm0.12$ \\
\end{tabular}
\vspace*{0.3cm}
\caption{Parameters from the fit to the functions Eq.~\ref{eq:parameterA} and Eq.~\ref{eq:parameterb} to
estimate the normalization parameter $A$ used to get the total number of collisions with the central runaway 
star in a cluster depending on the initial stellar radii R$_{\rm star}$ (Eq.~\ref{eq:parameterA}). These 
parameters are also used to estimate the parameter $t_{\rm delay}$ which is related to the time when 
the rate of collisions with the central runaway star is maximum, which also depends on the initial 
radii R$_{\rm star}$ of the stars (Eq.~\ref{eq:parameterb}).}
\label{table2}
\end{table}

\begin{figure}[h]
\includegraphics[width=250px]{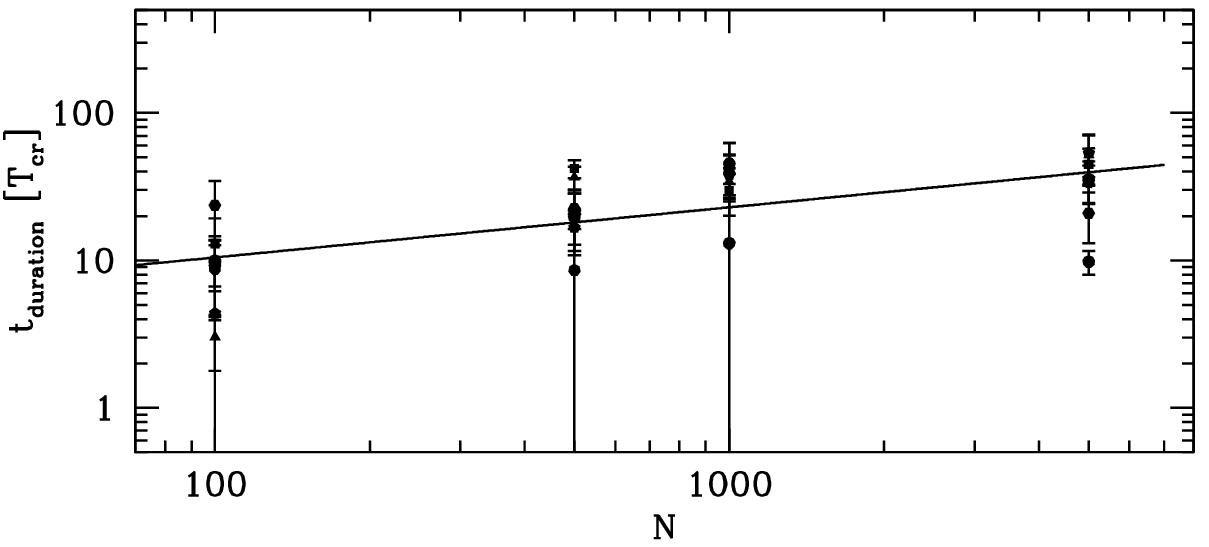}
\caption{Parameter $t_{\rm duration}$ of the gaussian fit as function of the number
of stars $N$. We assumed that $t_{\rm duration}$ only depends on $N$ and the fit is described in Eq.~\ref{eq:CvsN}.}
\label{gaussianparametersc}
\end{figure}

There is no clear dependence of the duration of the runaway growth on stellar radii
(see bottom panel of Fig.~\ref{gaussianparameters}). For that reason we explore the dependence of this 
parameter on the number of stars $N$. We find an exponential dependence for the duration of the 
runaway growth as function of the number of stars as shown in Fig.~\ref{gaussianparametersc}.
The relation between $t_{\rm duration}$ and $N$ is described as:\\

\begin{eqnarray}
\label{eq:CvsN}
\log(t_{\rm duration}) & = & 0.34 \pm 0.08 \log{N} + 0.34 \pm 0.24 .
\end{eqnarray}

As we find no clear relation between $t_{\rm duration}$ and R$_{\rm star}$ we
calculate this parameter only as function of $N$ from Eq.~\ref{eq:CvsN} and assume that 
$t_{\rm duration}$ is constant for all stellar radii. On the other hand, we have found clear relations for 
$A($R$_{\rm star})$ and $t_{\rm delay}($R$_{\rm star})$, however, these relations
depend also on $N$ (see Fig.~\ref{gaussianparameters}). In order
to get an approximate value for the number of collisions experienced by the runaway star in a star cluster consisting of
$N$ particles with a radii of R$_{\rm star}$ we need to take into account the dependence of $A$ and $t_{\rm delay}$ on $N$.\\
Equations ~\ref{eq:parameterA} and ~\ref{eq:parameterb}
show the dependence of $A$ and $t_{\rm delay}$ on R$_{\rm star}$. The slope
of these relations depends on the number of stars $N$ as described in Table ~\ref{table2}.
In order to properly account for the dependence on $N$ when calculating the number of collisions 
experienced by the runaway star, we fit a line for the parameters
$\alpha$, $\beta$, $\gamma$ and $\delta$ (related to the estimation
of $A$ and $t_{\rm delay}$) as function of $N$. The relations we found are described as:\\

\begin{eqnarray}
\label{eq:alpha}
\alpha &=& \ \ 0.16 \pm 0.09 \log{N} + 0.06 \pm 0.27 , \\
\label{eq:beta}
\beta  &=& -0.05 \pm 0.31 \log{N} - 1.43 \pm 0.90 , \\
\label{eq:gamma}
\gamma &=& -0.21 \pm 0.05 \log{N} + 0.30 \pm 0.13 , \\
\label{eq:delta}
\delta &=& \ \ 1.09 \pm 0.01  \log{N} - 0.79 \pm 0.04 .
\end{eqnarray}

If we then combine equations ~\ref{eq:parameterA}, ~\ref{eq:parameterb}, and \ref{eq:alpha} - 
\ref{eq:delta} we find $A$ and $t_{\rm delay}$ as a function of $N$  and R$_{\rm star}$ as
described in Eq.~\ref{eq:AonRandN} \& Eq.~\ref{eq:bonRandN}:\\

\begin{eqnarray}
\label{eq:AonRandN}
\log(A) &=& \left[ 0.16 \log(R_{\rm star}) - 0.05\right]\log(N) + \nonumber \\
& & 0.06\log(R_{\rm star}) -1.43 , \\
\label{eq:bonRandN}
\log(t_{\rm delay}) &=& \left[ -0.21 \log(R_{\rm star}) + 1.09\right]\log(N) + \nonumber \\ 
& & 0.30\log(R_{\rm star}) -0.79 .
\end{eqnarray}

Finally, the number of collisions experienced by the runaway star in a cluster of $N$ stars 
with radii of $R_{\rm star}$ between times $t_1$ and $t_2$ is given by:\\

\begin{eqnarray}
\label{eq:final}
 N_{\rm col} &=& A \int^{t_2}_{t_1} \exp \left\lbrace \frac{-(t-t_{\rm delay})^2}{2t_{\rm duration}^2} \right\rbrace dt ,
\end{eqnarray}
where $A$ is calculated from Eq.~\ref{eq:AonRandN}, $t_{\rm delay}$ is calculated from 
Eq.~\ref{eq:bonRandN} and $t_{\rm duration}$ is calculated from Eq.~\ref{eq:CvsN}. Time is expressed
in units of the cluster's crossing time.

\begin{figure}[h]
\includegraphics[width=250px]{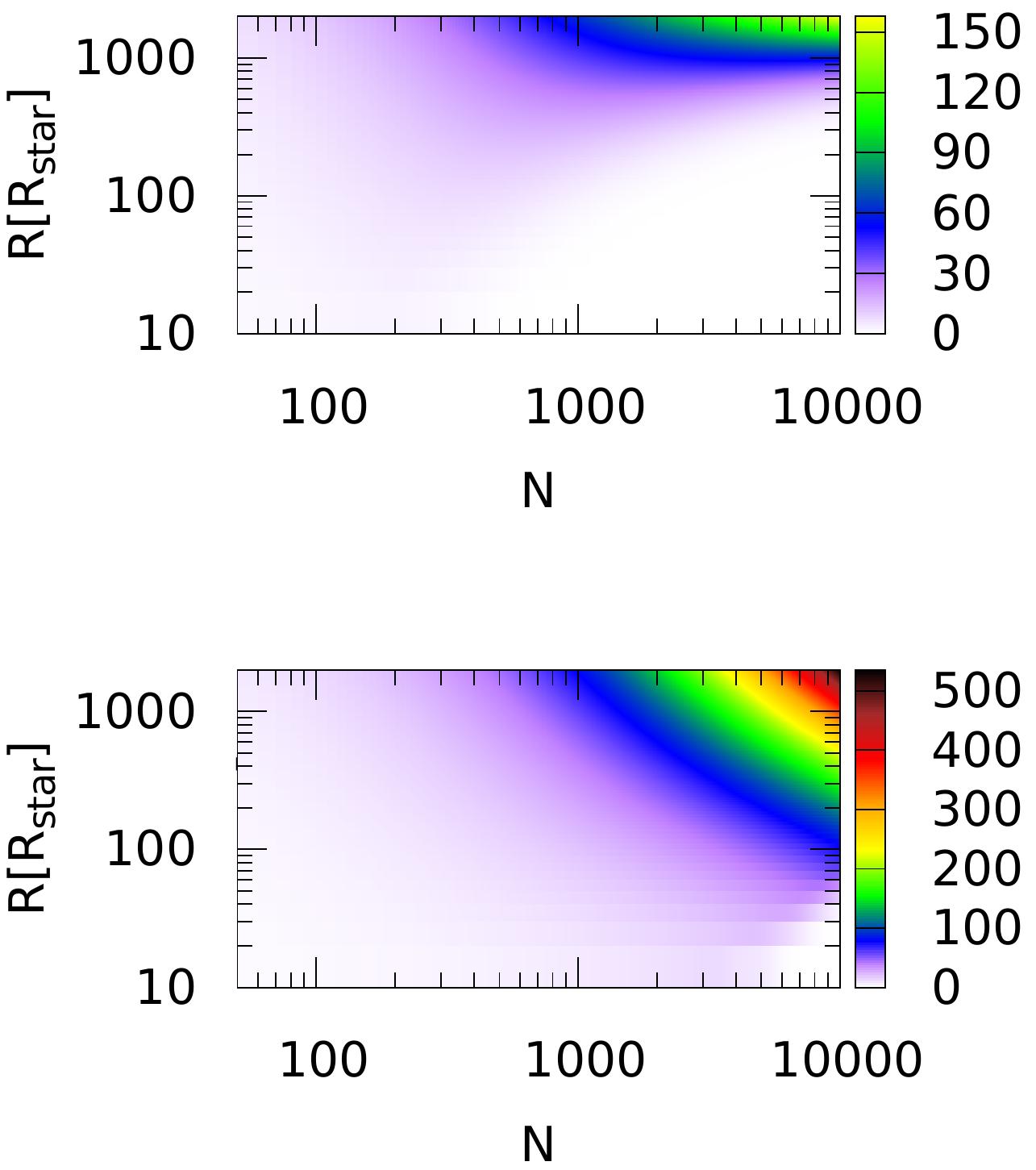}
\caption{Total number of collisions calculated via Eq.~(\ref{eq:final}) (displayed through color) with 
the central object after 1 Myr (Top panel) and after 10 Myr (Bottom panel) for our cluster model depending on the
number of stars and their initial radii.}
\label{fig:surfaceplot}
\end{figure}

We solve Eq.~\ref{eq:final} using the relations
found for $A$, $t_{\rm delay}$ and $t_{\rm duraion}$ as function of $N$ and $R_{\rm star}$
for $50 \leq N \leq 10000$ and $10 \leq R_{\rm star} \leq 2000$. We show
the expected number of collisions with the central object
after 1 Myr and 10 Myr in Fig.~\ref{fig:surfaceplot}. If collisions need to take place within 
10$^6$ yrs, due to the lifetime of the most massive star, the strongest collisional contributions 
may occur for clusters with about 1000 stars, as otherwise the characteristic timescale for 
the collisions to set in  becomes too long. In case of stellar radii of ~100~R$_{\odot}$, the enhancement 
by collisions corresponds to a factor of a few, which increases strongly for larger radii. In case 
the time available to form very massive objects is 10 million years, then the number of collisions increases
as a function of $N$ until a value of 5000 or higher, depending on the stellar radii. In such cases, 
relevant enhancements can be found even for stellar radii of 20~R$_{\odot}$, and potentially much larger 
for larger radii. This case is therefore the most promising case for the formation of very massive objects. It basically 
requires stars with less than 20~ M$_\odot$, as their lifetime is then longer than 10 Myr \citep{Schaerer2002}. Our simulations
show that in stellar systems containing equal mass stars with masses M $\leq$ 20 M$_\odot$, the expected enhancenment factor 
within 10 Myr ranges from 11-50 times the initial stellar mass M$_{\rm ini}$, considering
an intial stellar radius of 100 R$_{\odot}$ and depending on the number of stars.

In summary, we have provided a fit for $N_{\rm col}$ which depends on the typical duration of the collision phase 
$t_{\rm duration}$, the delay time until collisions occur $t_{\rm delay}$ as well as the overall normalization $A$. 
We found that $A$ and $t_{\rm duration}$ predominantly depend on the number of stars $N$, while $t_{\rm delay}$ 
depends both on $N$ and $R_{\rm star}$. We provide fits for the functional dependence of these parameters, and also 
demonstrate how the number of collisions that has occured after 1 and 10 million years depends both on $N$ and 
$R_{\rm star}$. We find that, as the delay time increases with $N$, the number of collisions within 10$^6$ yrs 
does not strongly increase with $N$ at fixed $R_{\rm star}$, but it does when considering a time of 10 Myr. 

\subsection{Ejections from the cluster}
\begin{figure}[h]
\includegraphics[width=250px]{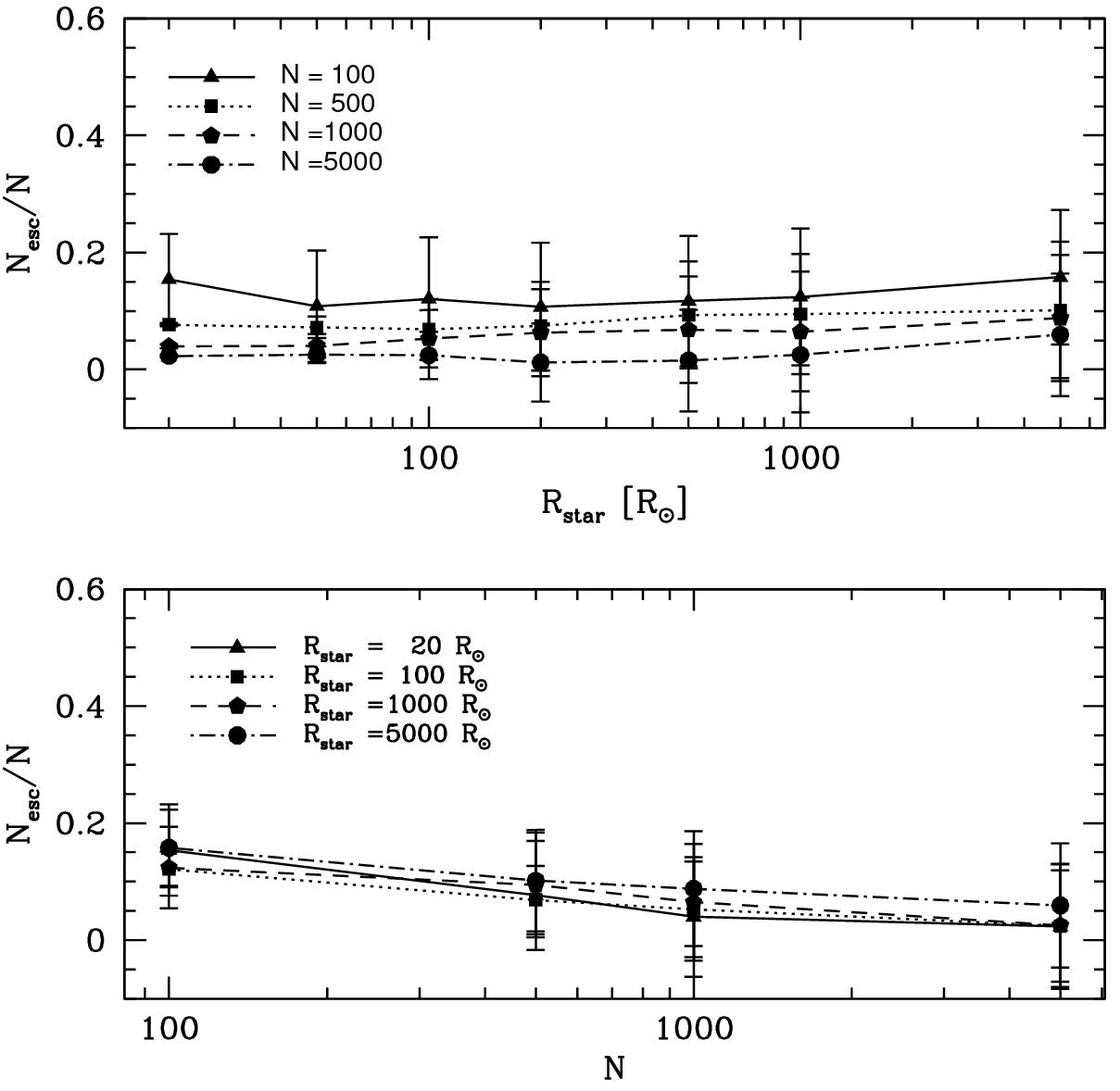}
\caption{Top panel: Fraction of ejections (number of ejected stars divided by the initial number of stars) as 
function of the initial stellar radius. Bottom panel: Fraction of ejections (number of ejected stars divided by the 
initial number of stars) as function of the initial number of stars.}
\label{ejections}
\end{figure}

We also investigate the number of ejection events, which are shown in Fig.~\ref{ejections} both as a 
function of stellar radius and as a function of the number of stars. A star is considered unbound once
its distance is 20 R$_{\rm vir}$. For all the clusters modeled in this work, the virial radius is
R$_{\rm vir}=0.14$ pc, thus a star has escaped the cluster when it is at 2.8 pc away from the cluster 
center. The fraction of escaped stars seems to vary from a few up to about $20\%$. While there is a large 
scatter for a given stellar radius, the escape fraction seems independent of that quantity, while it appears
to slightly decrease with the number of stars. This may particularly reflect that clusters with low 
numbers of stars of order $100$ dissolve more easily and do not provide a well-sampled statistical 
distribution. We have further checked that the velocity of the escaping stars is independent of stellar 
radius and number of stars, and corresponds to the expected escape velocity given the mass and radius of the cluster.

While stellar ejections have not yet been systematically explored in Pop.~III clusters, their potential role has 
been examined in different contexts. For instance, \citet{Pflamm2010} investigated the combined effect of massive binary
ejections from star clusters, including a second acceleration of a massive star during a subsequent supernova. The latter 
may potentially help to understand the observed fraction of isolated O-star formation candidates. \citet{Oh2012} explored 
how the ejection of massive stars affects the relation between maximum stellar mass and cluster mass. They find that 
lower mass clusters do not shoot out their heaviest star, while it may occur in more massive ones. These results have 
been further refined by \citet{Oh2015} and \citet{Oh2016}, showing that star clusters of $400$~M$_\odot$ are likely the 
dominant sources for O stars.

\subsection{Uncertainties and neglected processes}\label{uncertainties}
In this investigation, we have not included the effect of a gaseous potential or modeled the effect of the gas
in explicit terms. To zero order, if the latter is dominating the potential, it will primarily increase the 
velocity dispersion and thereby decrease the crossing time of the cluster, allowing for a larger number of 
crossing times in a given physical time. In addition, the gas may provide dynamical friction, which can 
potentially favor the probability of collisions. The effects of dark matter are also neglected given that at 
the point when the star clusters are formed, the gravitational potential is dominated by the baryons \citep{Abel02,Bromm03}.

Related to this, the accretion onto the protostars was not explicitly considered, and their masses were taken 
to be constant. In case that accretion contributes to the mass growth, it may again favor collisions and induce
mergers of binaries that otherwise would be stable. 

A further simplifying assumption employed here was that we did not consider a full initial mass function (IMF), 
but assumed initially uniform stellar masses. The latter has the effect to suppress three-body ejections, 
which may otherwise be more frequent and preferably eject low-mass stars. In the presence of a logarithmically
flat IMF, as it is often assumed in the context of primordial star formation \citep{Greif12, Hartwig16}, we 
do not expect this to be a major problem, though it can possible reduce the mass of the most massive object 
by a certain degree. We note here also that the simulations by \citet{Sakurai17} included a full IMF, which 
however has not prevented the formation of a $1000$~M$_\odot$ object.

In case a full IMF is considered, the next step is then clearly to take the dependence of the mass on stellar 
radius into account, as well as the evolution of the stellar radii over time. In this sense, our current 
simulations represent an effective model where an average stellar radius is adopted over the period of time 
considered. This effective stellar radius should correspond to the typical stellar radius at the time when the
majority of collisions is expected to occur. In addition, also the "hit-and-stick" assumption employed here is valid for:
\begin{eqnarray}
\label{eq:hitstick}
\sigma^{2}_{\rm stars} < G\frac{M_{\rm star}}{R_{\rm star}} .
\end{eqnarray}
The velocity dispersion of the stars in all our simulations
is 12.39 km s$^{-1}$, thus the hit-and-stick approximation should be valid for all our models except 
when R$_{\rm star} = 5000$ R$_{\odot}$. However this approximation still needs to be tested in future 
simulations, both to investigate how much material needs to stick for efficient growth still to occur, 
but also specific simulations exploring stellar collisions and merger processes would be valuable for a 
better understanding of possible limitations. 

While clearly the processes mentioned here deserve further investigations, we will show below that the 
results obtained above clearly suggest the potential relevance of collisions in primordial clusters.

\section{Implications for primordial clusters}\label{application}
In the following, we explore the implications of our results with respect to primordial star 
clusters. We distinguish here in particular the case of standard Pop.~III clusters as expected in a 
typical minihalo with about $10^6$~M$_\odot$, and a more massive atomic cooling halo with about $10^8$~M$_\odot$.

\subsection{Standard Pop. III clusters (minihalos)}
For a typical Pop.~III star cluster, we assume here a mass of $1000$~M$_\odot$, consistent 
with a baryon fraction of about $10\%$ and a star formation efficiency of order $1\%$ in a $10^6$~M$_\odot$ 
minihalo. We adopt here a radius of the cluster of order $0.1$~pc, consistent with results from simulations 
and semi-analytic models \citep{Clark11a, Clark11b, Greif11, Greif12,Latif13PopIII,Latif15Disk}. We take a
stellar radius of $100$~R$_\odot$, which is characteristic for primordial protostars with accretion rates 
of the order $10^{-3}$~M$_\odot$~yr$^{-1}$ \citep{Hosokawa12}. The crossing time of the cluster then 
corresponds to $0.071$~Myr. The number of stars that can be expected in such a cluster is uncertain, but we 
adopt here an estimate of about $100$.

Using the relations we found, we expect a total of about $4$ collisions to occur within $1$ million
years. This is unchanged even if we assume a lifetime of 10~Myr, as the runaway growth occurs within 
1 Myr. We expect the lifetime of a massive primordial star to be in between this range, depending on precise
mass, amount of rotation and the effects the collisions may have on the stellar evolution 
\citep[e.g.][]{Maeder12}. We therefore find that a moderate enhancement can be achieved within a normal cluster.

We note that the values given here are the expected mean number of collisions. Individual clusters can deviate from 
these both towards lower and higher fractions of collisions, including potentially clusters with zero collisions. Especially 
for typical Pop.~III clusters where the number of collisions is low, the uncertainty in the collision fraction can be expected 
to be comparable to the mean value.

\subsection{Massive primordial clusters (atomic cooling halos)}
As a next step, we address now the potential impact of collisions in a more massive atomic 
cooling halo. Under the right conditions, in particular if the cooling on larger scales is regulated by
atomic hydrogen \citep[e.g.][]{Latif14Star}, a rather massive cluster of  $10^4$~M$_\odot$ can form,
which will be exposed to larger accretion rates of the order $10^{-1}$~M$_\odot$~yr$^{-1}$. We assume 
that the cluster consists of an initial amount of $1000$ stars, and that the stellar radii are somewhat
enhanced compared to the standard Pop.~III cluster due to the higher accretion, with a typical value of 
about $300$~R$_\odot$. The crossing time in the cluster is then $0.023$~Myr.

Using the relations derived above, we expect about 15 collisions in 1 million years, and about 32 
within 10 million years with a single runaway star. We again expect the realistic lifetime of the 
resulting massive star to be in between these extreme cases. In the case of an atomic cooling halo, 
we thus conclude that a more considerable enhancement of the mass is possible as a result of stellar collisions.

Also here, the reported values correspond to a mean, and there can be deviations to lower and higher collision 
fractions. We however expect the number of collisions to remain within the same order of magnitude.

\section{Discussion and conclusions}\label{discussion}
In this paper, we have provided a detailed parameter study showing how the number of collisions and 
the growth of the mass of the most massive object depends on the properties of the stellar cluster,
particularly the number of stars and the stellar radii. We have further quantified their time evolution,
providing the time when most collisions occur, the duration of the peak collision period as well as the
overall normalization via a Gaussian fit. We provide fits and scaling relations on the general results
that we found, concerning the characteristic time delay until the maximum of collisions occurs, the duration 
of the period of collisions as well as the normalization that determines the number of collisions during 
that period. We have applied our results both to the formation of a standard Pop.~III star cluster, finding a 
moderate enhancement of the mass of the most massive star by a factor of a few, as well as to a massive 
Pop.~III cluster in a larger atomic cooling halo, finding a potential enhancement by a factor $15-32$.

Assuming a typical protostellar mass of $20$~M$_\odot$, this would as a result imply resulting black hole masses of 
up to $600$~M$_\odot$, compatible with simulation results by \citet{Katz15} and \citet{Sakurai17}. We do however note that
especially in the massive atomic cooling halos, the properties of the Pop.~III clusters are not very well known, and it 
may be conceivable that a more massive object can form depending on mass and compactness of the clusters. Taking the 
results obtained here as a conservative estimate (as we neglected the gas-phase processes in this investigation), the 
resulting black hole masses would correspond to about $0.1-10\%$ of black hole masses formed via direct collapse 
\citep{Koushiappas04, Lodato07, Latif13,Schleicher13, Ferrara14}.

We find both results potentially relevant. The rather modest increase of the mass of the most massive star 
in standard Pop.~III clusters could on the one hand be relevant in case it then falls into the range of 
$130-260$~M$_\odot$ where pair-instability supernovae are expected to occur \citep{Heger02}. On the other 
hand, as no pair-instability abundance patterns have so far been found in extremely metal poor stars 
\citep{Frebel15}, such a small factor may also explain why pair-instability supernovae did not generally occur,
or have been too rare events to be found. While it is likely too early to draw definite conclusions, 
we expect that ongoing efforts through stellar archeology as well as the upcoming 
JWST\footnote{Webpage JWST: https://www.jwst.nasa.gov/} will shed light on this issue.

In massive Pop.~III clusters, our finding that the mass of the most massive object increases can be 
expected to be highly relevant. This is particular because the formation of massive seeds via direct 
collapse still seems highly challenging, requiring both very low metallicities below 
$10^{-5}$~Z$_\odot$ \citep{Omukai08,Latif15dust}, as well as very strong radiation backgrounds to 
dissociate the molecular hydrogen \citep{Latif15X}. Such a scenario may perhaps work as a rare 
event \citep[e.g.][]{Habouzit16}. A more recent theory suggesting the existence of
mirror dark matter, that is, photons and neutrinos with the same characteristics as in the standard
model of particles but interacting only via gravity, could explain the formation of 10$^4$-10$^5$ M$_{\odot}$
black holes at very high redshift, becoming SMBHs at redshift 7 wthout the need to accrete at
super-Eddington rates \citep{DAmico2018}. If collisions are taken into account, on the other hand, the formation 
of massive objects may become feasible under a broader range of circumstances.

However, even if fragmentation occurs, it is very likely that a relevant fraction of the mass nevertheless
ends up within the most massive object, as we showed here. In these calculations, we did not yet take into
account dissipative effects such as dynamical friction, or the effects resulting from ongoing accretion
onto the stars. We therefore expect that particularly within the actively accreting clusters, the number
of collisions may be further enhanced, even though this could be partly balanced from the effect of 
having a more realistic initial mass function (IMF), that could enhance the stellar ejections. While it
is clear that further investigations will be required, the present study shows clearly the prospect 
of forming massive objects via stellar collisions.

\begin{acknowledgements}
BR thanks Sverre Aarseth for his help with the code NBODY6. BR also thanks A. Alarcon Jara and D.R.
Matus Carrillo for useful discussion during the realization of this work. RSK thanks B. Agarwal,
L. Haemmerl\'e, A. Heger, D. Whalen, and T. Woods for stimulating discussions about 
massive and supermassive Pop. III stars. DRGS and BR thank for 
funding through Fondecyt regular (project code 1161247). BR thanks Conicyt for 
financial support (CONICYT-PFCHA/Mag\'isterNacional/2017-22171385). DRGS and MF acknowledge 
funding through the ''Concurso Proyectos Internacionales de Investigaci\'on, 
Convocatoria 2015'' (project code PII20150171) and the BASAL Centro de Astrof\'isica y 
Tecnolog\'ias Afines (CATA) PFB-06/2007. DRGS further is grateful for funding via ALMA-Conicyt
(project code 31160001), Quimal (project number QUIMAL170001) and Anillo (project number ACT172033). RSK acknowledges 
financial support from the Deutsche Forschungsgemeinschaft via SFB 881, ``The Milky Way System'' (sub-projects B1, B2 and B8) 
and SPP 1573, ``Physics of the Interstellar Medium''. He also thanks for support from the 
European Research Council via the ERC Advanced Grant ``STARLIGHT: Formation of the First Stars'' (project number 339177). 
TB acknowledges support from Funda\c{c}\~ ao para a Ci\^ encia e a Tecnologia (grant SFRH/BPD/122325/2016), and support from 
Center for Research \& Development in Mathematics and Applications (CIDMA) (strategic project UID/MAT/04106/2013), and 
from ENGAGE SKA, POCI-01-0145-FEDER-022217, funded by COMPETE 2020 and FCT, Portugal.
\end{acknowledgements}



\begin{thebibliography}{74}
\expandafter\ifx\csname natexlab\endcsname\relax\def\natexlab#1{#1}\fi

\bibitem[{{Aarseth}(2000)}]{Aarseth2000}
{Aarseth}, S.~J. 2000, in The Chaotic Universe, ed. V.~G. {Gurzadyan} \&
  R.~{Ruffini}, 286--287

\bibitem[{{Abel} {et~al.}(2002){Abel}, {Bryan}, \& {Norman}}]{Abel02}
{Abel}, T., {Bryan}, G.~L., \& {Norman}, M.~L. 2002, Science, 295, 93

\bibitem[{{Ahmad} \& {Cohen}(1973)}]{Ahmad1973}
{Ahmad}, A. \& {Cohen}, L. 1973, Journal of Computational Physics, 12, 389

\bibitem[{{Baumgardt} \& {Klessen}(2011)}]{Baumgardt11}
{Baumgardt}, H. \& {Klessen}, R.~S. 2011, \mnras, 413, 1810

\bibitem[{{Bodenheimer} \& {Sweigart}(1968)}]{Bodenheimer68}
{Bodenheimer}, P. \& {Sweigart}, A. 1968, \apj, 152, 515

\bibitem[{{Bonnell} {et~al.}(1998){Bonnell}, {Bate}, \&
  {Zinnecker}}]{Bonnell98}
{Bonnell}, I.~A., {Bate}, M.~R., \& {Zinnecker}, H. 1998, \mnras, 298, 93

\bibitem[{{Bovino} {et~al.}(2016){Bovino}, {Grassi}, {Schleicher}, \&
  {Banerjee}}]{Bovino16}
{Bovino}, S., {Grassi}, T., {Schleicher}, D.~R.~G., \& {Banerjee}, R. 2016,
  \apj, 832, 154

\bibitem[{{Bromm} \& {Loeb}(2003)}]{Bromm03}
{Bromm}, V. \& {Loeb}, A. 2003, ApJ, 596, 34

\bibitem[{{Clark} {et~al.}(2011{\natexlab{a}}){Clark}, {Glover}, {Klessen}, \&
  {Bromm}}]{Clark11b}
{Clark}, P.~C., {Glover}, S.~C.~O., {Klessen}, R.~S., \& {Bromm}, V.
  2011{\natexlab{a}}, \apj, 727, 110

\bibitem[{{Clark} {et~al.}(2011{\natexlab{b}}){Clark}, {Glover}, {Smith},
  {Greif}, {Klessen}, \& {Bromm}}]{Clark11a}
{Clark}, P.~C., {Glover}, S.~C.~O., {Smith}, R.~J., {et~al.}
  2011{\natexlab{b}}, Science, 331, 1040

\bibitem[{{Clarke} \& {Bonnell}(2008)}]{Clarke08}
{Clarke}, C.~J. \& {Bonnell}, I.~A. 2008, \mnras, 388, 1171

\bibitem[{{D'Amico} {et~al.}(2018){D'Amico}, {Panci}, {Lupi}, {Bovino}, \&
  {Silk}}]{DAmico2018}
{D'Amico}, G., {Panci}, P., {Lupi}, A., {Bovino}, S., \& {Silk}, J. 2018,
  \mnras, 473, 328

\bibitem[{{Devecchi} {et~al.}(2010){Devecchi}, {Volonteri}, {Colpi}, \&
  {Haardt}}]{Devecchi10}
{Devecchi}, B., {Volonteri}, M., {Colpi}, M., \& {Haardt}, F. 2010, \mnras,
  409, 1057

\bibitem[{{Devecchi} {et~al.}(2012){Devecchi}, {Volonteri}, {Rossi}, {Colpi},
  \& {Portegies Zwart}}]{Devecchi12}
{Devecchi}, B., {Volonteri}, M., {Rossi}, E.~M., {Colpi}, M., \& {Portegies
  Zwart}, S. 2012, \mnras, 421, 1465

\bibitem[{{Dopcke} {et~al.}(2011){Dopcke}, {Glover}, {Clark}, \&
  {Klessen}}]{Dopcke11}
{Dopcke}, G., {Glover}, S.~C.~O., {Clark}, P.~C., \& {Klessen}, R.~S. 2011,
  \apjl, 729, L3

\bibitem[{{Dopcke} {et~al.}(2013){Dopcke}, {Glover}, {Clark}, \&
  {Klessen}}]{Dopcke13}
{Dopcke}, G., {Glover}, S.~C.~O., {Clark}, P.~C., \& {Klessen}, R.~S. 2013,
  \apj, 766, 103

\bibitem[{{Ferrara} {et~al.}(2014){Ferrara}, {Salvadori}, {Yue}, \&
  {Schleicher}}]{Ferrara14}
{Ferrara}, A., {Salvadori}, S., {Yue}, B., \& {Schleicher}, D. 2014, \mnras,
  443, 2410

\bibitem[{{Frebel} \& {Norris}(2015)}]{Frebel15}
{Frebel}, A. \& {Norris}, J.~E. 2015, \araa, 53, 631

\bibitem[{{Fujii} \& {Portegies Zwart}(2013)}]{Fuji13}
{Fujii}, M.~S. \& {Portegies Zwart}, S. 2013, \mnras, 430, 1018

\bibitem[{{Greif} {et~al.}(2012){Greif}, {Bromm}, {Clark}, {Glover}, {Smith},
  {Klessen}, {Yoshida}, \& {Springel}}]{Greif12}
{Greif}, T.~H., {Bromm}, V., {Clark}, P.~C., {et~al.} 2012, \mnras, 424, 399

\bibitem[{{Greif} {et~al.}(2011){Greif}, {Springel}, {White}, {Glover},
  {Clark}, {Smith}, {Klessen}, \& {Bromm}}]{Greif11}
{Greif}, T.~H., {Springel}, V., {White}, S.~D.~M., {et~al.} 2011, \apj, 737, 75

\bibitem[{{Habouzit} {et~al.}(2016){Habouzit}, {Volonteri}, {Latif}, {Dubois},
  \& {Peirani}}]{Habouzit16}
{Habouzit}, M., {Volonteri}, M., {Latif}, M., {Dubois}, Y., \& {Peirani}, S.
  2016, \mnras, 463, 529

\bibitem[{{Haemmerl{\'e}} {et~al.}(2016){Haemmerl{\'e}}, {Eggenberger},
  {Meynet}, {Maeder}, \& {Charbonnel}}]{Haemmerle16}
{Haemmerl{\'e}}, L., {Eggenberger}, P., {Meynet}, G., {Maeder}, A., \&
  {Charbonnel}, C. 2016, \aap, 585, A65

\bibitem[{{Haemmerl{\'e}} {et~al.}(2017){Haemmerl{\'e}}, {Woods}, {Klessen},
  {Heger}, \& {Whalen}}]{Haemmerle17}
{Haemmerl{\'e}}, L., {Woods}, T.~E., {Klessen}, R.~S., {Heger}, A., \&
  {Whalen}, D.~J. 2017, ArXiv e-prints [\eprint[arXiv]{1705.09301}]

\bibitem[{{Hartwig} {et~al.}(2016){Hartwig}, {Volonteri}, {Bromm}, {Klessen},
  {Barausse}, {Magg}, \& {Stacy}}]{Hartwig16}
{Hartwig}, T., {Volonteri}, M., {Bromm}, V., {et~al.} 2016, \mnras, 460, L74

\bibitem[{{Heger} \& {Woosley}(2002)}]{Heger02}
{Heger}, A. \& {Woosley}, S.~E. 2002, ApJ, 567, 532

\bibitem[{{Hosokawa} {et~al.}(2012){Hosokawa}, {Omukai}, \&
  {Yorke}}]{Hosokawa12}
{Hosokawa}, T., {Omukai}, K., \& {Yorke}, H.~W. 2012, \apj, 756, 93

\bibitem[{{Hosokawa} {et~al.}(2013){Hosokawa}, {Yorke}, {Inayoshi}, {Omukai},
  \& {Yoshida}}]{Hosokawa13}
{Hosokawa}, T., {Yorke}, H.~W., {Inayoshi}, K., {Omukai}, K., \& {Yoshida}, N.
  2013, \apj, 778, 178

\bibitem[{{Hurley} {et~al.}(2000){Hurley}, {Pols}, \& {Tout}}]{Hurley2000}
{Hurley}, J.~R., {Pols}, O.~R., \& {Tout}, C.~A. 2000, \mnras, 315, 543

\bibitem[{{Katz} {et~al.}(2015){Katz}, {Sijacki}, \& {Haehnelt}}]{Katz15}
{Katz}, H., {Sijacki}, D., \& {Haehnelt}, M.~G. 2015, \mnras, 451, 2352

\bibitem[{{Keto} \& {Klaassen}(2008)}]{Keto08}
{Keto}, E. \& {Klaassen}, P. 2008, \apjl, 678, L109

\bibitem[{{Klessen} {et~al.}(2012){Klessen}, {Glover}, \& {Clark}}]{Klessen12}
{Klessen}, R.~S., {Glover}, S.~C.~O., \& {Clark}, P.~C. 2012, \mnras, 421, 3217

\bibitem[{{Koushiappas} {et~al.}(2004){Koushiappas}, {Bullock}, \&
  {Dekel}}]{Koushiappas04}
{Koushiappas}, S.~M., {Bullock}, J.~S., \& {Dekel}, A. 2004, \mnras, 354, 292

\bibitem[{{Krumholz} {et~al.}(2009){Krumholz}, {Klein}, {McKee}, {Offner}, \&
  {Cunningham}}]{Krumholz09}
{Krumholz}, M.~R., {Klein}, R.~I., {McKee}, C.~F., {Offner}, S.~S.~R., \&
  {Cunningham}, A.~J. 2009, Science, 323, 754

\bibitem[{{Kustaanheimo} \& {Stiefel}(1965)}]{Kustaanheimo65}
{Kustaanheimo}, P. \& {Stiefel}, E. 1965, Journal f\"ur die Reine und
  Angewandte Mathematik, 218, 204

\bibitem[{{Larson}(1969)}]{Larson69}
{Larson}, R.~B. 1969, \mnras, 145, 271

\bibitem[{{Latif} {et~al.}(2015{\natexlab{a}}){Latif}, {Bovino}, {Grassi},
  {Schleicher}, \& {Spaans}}]{Latif15X}
{Latif}, M.~A., {Bovino}, S., {Grassi}, T., {Schleicher}, D.~R.~G., \&
  {Spaans}, M. 2015{\natexlab{a}}, \mnras, 446, 3163

\bibitem[{{Latif} {et~al.}(2015{\natexlab{b}}){Latif}, {Omukai}, {Habouzit},
  {Schleicher}, \& {Volonteri}}]{Latif15dust}
{Latif}, M.~A., {Omukai}, K., {Habouzit}, M., {Schleicher}, D.~R.~G., \&
  {Volonteri}, M. 2015{\natexlab{b}}, MNRAS, submitted (ArXiv e-prints
  1509.07034) [\eprint[arXiv]{1509.07034}]

\bibitem[{{Latif} {et~al.}(2016){Latif}, {Omukai}, {Habouzit}, {Schleicher}, \&
  {Volonteri}}]{Latif16}
{Latif}, M.~A., {Omukai}, K., {Habouzit}, M., {Schleicher}, D.~R.~G., \&
  {Volonteri}, M. 2016, \apj, 823, 40

\bibitem[{{Latif} \& {Schleicher}(2015)}]{Latif15Disk}
{Latif}, M.~A. \& {Schleicher}, D.~R.~G. 2015, \mnras, 449, 77

\bibitem[{{Latif} {et~al.}(2014){Latif}, {Schleicher}, {Bovino}, {Grassi}, \&
  {Spaans}}]{Latif14Star}
{Latif}, M.~A., {Schleicher}, D.~R.~G., {Bovino}, S., {Grassi}, T., \&
  {Spaans}, M. 2014, \apj, 792, 78

\bibitem[{{Latif} {et~al.}(2013{\natexlab{a}}){Latif}, {Schleicher}, {Schmidt},
  \& {Niemeyer}}]{Latif13PopIII}
{Latif}, M.~A., {Schleicher}, D.~R.~G., {Schmidt}, W., \& {Niemeyer}, J.
  2013{\natexlab{a}}, \apjl, 772, L3

\bibitem[{{Latif} {et~al.}(2013{\natexlab{b}}){Latif}, {Schleicher}, {Schmidt},
  \& {Niemeyer}}]{Latif13}
{Latif}, M.~A., {Schleicher}, D.~R.~G., {Schmidt}, W., \& {Niemeyer}, J.~C.
  2013{\natexlab{b}}, \mnras, 436, 2989

\bibitem[{{Lodato} \& {Natarajan}(2007)}]{Lodato07}
{Lodato}, G. \& {Natarajan}, P. 2007, \mnras, 377, L64

\bibitem[{{Lupi} {et~al.}(2014){Lupi}, {Colpi}, {Devecchi}, {Galanti}, \&
  {Volonteri}}]{Lupi2014}
{Lupi}, A., {Colpi}, M., {Devecchi}, B., {Galanti}, G., \& {Volonteri}, M.
  2014, \mnras, 442, 3616

\bibitem[{{Maeder} \& {Meynet}(2012)}]{Maeder12}
{Maeder}, A. \& {Meynet}, G. 2012, Reviews of Modern Physics, 84, 25

\bibitem[{{Moeckel} \& {Clarke}(2011)}]{Moeckel11}
{Moeckel}, N. \& {Clarke}, C.~J. 2011, \mnras, 410, 2799

\bibitem[{{Oh} \& {Kroupa}(2012{\natexlab{a}})}]{Oh12}
{Oh}, S. \& {Kroupa}, P. 2012{\natexlab{a}}, \mnras, 424, 65

\bibitem[{{Oh} \& {Kroupa}(2012{\natexlab{b}})}]{Oh2012}
{Oh}, S. \& {Kroupa}, P. 2012{\natexlab{b}}, \mnras, 424, 65

\bibitem[{{Oh} \& {Kroupa}(2016)}]{Oh2016}
{Oh}, S. \& {Kroupa}, P. 2016, \aap, 590, A107

\bibitem[{{Oh} {et~al.}(2015){Oh}, {Kroupa}, \& {Pflamm-Altenburg}}]{Oh2015}
{Oh}, S., {Kroupa}, P., \& {Pflamm-Altenburg}, J. 2015, \apj, 805, 92

\bibitem[{{Omukai} \& {Palla}(2001)}]{Omukai01}
{Omukai}, K. \& {Palla}, F. 2001, \apjl, 561, L55

\bibitem[{{Omukai} \& {Palla}(2003)}]{Omukai03}
{Omukai}, K. \& {Palla}, F. 2003, \apj, 589, 677

\bibitem[{{Omukai} {et~al.}(2008){Omukai}, {Schneider}, \& {Haiman}}]{Omukai08}
{Omukai}, K., {Schneider}, R., \& {Haiman}, Z. 2008, \apj, 686, 801

\bibitem[{{Omukai} {et~al.}(2005){Omukai}, {Tsuribe}, {Schneider}, \&
  {Ferrara}}]{Omukai05}
{Omukai}, K., {Tsuribe}, T., {Schneider}, R., \& {Ferrara}, A. 2005, \apj, 626,
  627

\bibitem[{{Palla} \& {Stahler}(1993)}]{Palla93}
{Palla}, F. \& {Stahler}, S.~W. 1993, \apj, 418, 414

\bibitem[{{Peters} {et~al.}(2011){Peters}, {Banerjee}, {Klessen}, \& {Mac
  Low}}]{Peters11}
{Peters}, T., {Banerjee}, R., {Klessen}, R.~S., \& {Mac Low}, M.-M. 2011, \apj,
  729, 72

\bibitem[{{Peters} {et~al.}(2010{\natexlab{a}}){Peters}, {Banerjee}, {Klessen},
  {Mac Low}, {Galv{\'a}n-Madrid}, \& {Keto}}]{Peters10a}
{Peters}, T., {Banerjee}, R., {Klessen}, R.~S., {et~al.} 2010{\natexlab{a}},
  \apj, 711, 1017

\bibitem[{{Peters} {et~al.}(2010{\natexlab{b}}){Peters}, {Klessen}, {Mac Low},
  \& {Banerjee}}]{Peters10b}
{Peters}, T., {Klessen}, R.~S., {Mac Low}, M.-M., \& {Banerjee}, R.
  2010{\natexlab{b}}, \apj, 725, 134

\bibitem[{{Pflamm-Altenburg} \& {Kroupa}(2010)}]{Pflamm2010}
{Pflamm-Altenburg}, J. \& {Kroupa}, P. 2010, \mnras, 404, 1564

\bibitem[{{Plummer}(1911)}]{Plummer1911}
{Plummer}, H.~C. 1911, \mnras, 71, 460

\bibitem[{{Rees}(1984)}]{Rees84}
{Rees}, M.~J. 1984, \araa, 22, 471

\bibitem[{{Sakurai} {et~al.}(2017){Sakurai}, {Yoshida}, {Fujii}, \&
  {Hirano}}]{Sakurai17}
{Sakurai}, Y., {Yoshida}, N., {Fujii}, M.~S., \& {Hirano}, S. 2017, ArXiv
  e-prints [\eprint[arXiv]{1704.06130}]

\bibitem[{{Schaerer}(2002)}]{Schaerer2002}
{Schaerer}, D. 2002, \aap, 382, 28

\bibitem[{{Schleicher} {et~al.}(2013){Schleicher}, {Palla}, {Ferrara}, {Galli},
  \& {Latif}}]{Schleicher13}
{Schleicher}, D.~R.~G., {Palla}, F., {Ferrara}, A., {Galli}, D., \& {Latif}, M.
  2013, \aap, 558, A59

\bibitem[{{Schneider} {et~al.}(2003){Schneider}, {Ferrara}, {Salvaterra},
  {Omukai}, \& {Bromm}}]{Schneider03}
{Schneider}, R., {Ferrara}, A., {Salvaterra}, R., {Omukai}, K., \& {Bromm}, V.
  2003, \nat, 422, 869

\bibitem[{{Schneider} {et~al.}(2006){Schneider}, {Omukai}, {Inoue}, \&
  {Ferrara}}]{Schneider06}
{Schneider}, R., {Omukai}, K., {Inoue}, A.~K., \& {Ferrara}, A. 2006, \mnras,
  369, 1437

\bibitem[{{Schneider} {et~al.}(2012){Schneider}, {Omukai}, {Limongi},
  {Ferrara}, {Salvaterra}, {Chieffi}, \& {Bianchi}}]{Schneider12}
{Schneider}, R., {Omukai}, K., {Limongi}, M., {et~al.} 2012, \mnras, 423, L60

\bibitem[{{Shu}(1977)}]{Shu77}
{Shu}, F.~H. 1977, \apj, 214, 488

\bibitem[{{Smith} {et~al.}(2011){Smith}, {Glover}, {Clark}, {Greif}, \&
  {Klessen}}]{Smith11}
{Smith}, R.~J., {Glover}, S.~C.~O., {Clark}, P.~C., {Greif}, T., \& {Klessen},
  R.~S. 2011, \mnras, 414, 3633

\bibitem[{{Smith} {et~al.}(2012){Smith}, {Hosokawa}, {Omukai}, {Glover}, \&
  {Klessen}}]{Smith12}
{Smith}, R.~J., {Hosokawa}, T., {Omukai}, K., {Glover}, S.~C.~O., \& {Klessen},
  R.~S. 2012, \mnras, 424, 457

\bibitem[{{Stahler} {et~al.}(1986){Stahler}, {Palla}, \&
  {Salpeter}}]{Stahler86}
{Stahler}, S.~W., {Palla}, F., \& {Salpeter}, E.~E. 1986, \apj, 302, 590

\bibitem[{{Woods} {et~al.}(2017){Woods}, {Heger}, {Whalen}, {Haemmerl{\'e}}, \&
  {Klessen}}]{Woods17}
{Woods}, T.~E., {Heger}, A., {Whalen}, D.~J., {Haemmerl{\'e}}, L., \&
  {Klessen}, R.~S. 2017, \apjl, 842, L6

\bibitem[{{Yorke} \& {Sonnhalter}(2002)}]{Yorke02}
{Yorke}, H.~W. \& {Sonnhalter}, C. 2002, \apj, 569, 846

\end{thebibliography}

\bibliographystyle{aa}

\end{document}